\documentstyle [12 pt]{article}
\begin{document}%

\title{An exact formula for general spectral
correlation function of random Hermitian matrices}
\author{ Yan V Fyodorov\footnote{on leave from Petersburg Nuclear Physics
Institute, Gatchina 188350, Russia} and Eugene Strahov
\footnote{Yan.Fyodorov@brunel.ac.uk,
Eugene.Strahov@brunel.ac.uk}\\
{\sl Department of Mathematical Sciences, Brunel University}\\
{\it Uxbridge, UB8 3PH, United Kingdom }} \maketitle

{\bf Published:} J.Phys.A:Math.Gen. v.36 (2003), 3203-3214

\begin{abstract}
We have found an exact formula expressing a general correlation
function containing both products and ratios of characteristic
polynomials of random Hermitian matrices. The answer is given in
the form of a determinant. An essential difference from the
previously studied correlation functions (of products only) is the
appearance of non-polynomial functions along with the orthogonal
polynomials. These non-polynomial functions are the Cauchy
transforms of the orthogonal polynomials. The result is valid for
arbitrary ensemble of $\beta=2$ symmetry class and generalizes
recent asymptotic formulas obtained for GUE and its chiral
counterpart by different methods.
\end{abstract}
\section{Introduction}
It is a classical result of analysis known from the nineteenth
century that orthogonal polynomials can be represented as
multi-variable integrals. For example, let $\pi_j(x)$ denote a
monic orthogonal polynomial of $j^{\mbox{th}}$ degree with respect
to the measure $d\mu(x)=\exp(-NV(x))dx$ on the real axis,
\begin{equation}
\int\pi_j(x)\pi_k(x)e^{-NV(x)}dx=c_jc_k\delta_{jk}
\end{equation}
and $\pi_{j}(x)=x^j+\mbox{lower}\;\mbox{degrees}$. Then for the
monic orthogonal polynomial $\pi_{N}(x)$ there exists an integral
representation (see Szeg$\ddot{\mbox{o}}$ \cite{szego})
\begin{equation}\label{IntRepOrthPol}
\pi_{N}(x)= \frac{1}{Z_N}\int\prod\limits_{j=1}^{N}(x
-x_j)\;e^{-N\sum\limits_{j=1}^{N}V(x_j)}\triangle^2(x_1,\ldots ,
x_N)dx_1\ldots x_N
\end{equation}
In the formula above $\triangle(x_1,\ldots ,x_N)$ stands for the
Vandermonde determinant, and the constant $Z_N$ is given by the
product $Z_N=N!\prod\limits_{j=0}^{N-1}c_j^2$.

The classical integral representation  of the monic orthogonal
polynomial $\pi_{N}(x)$ given in (\ref{IntRepOrthPol}) suggests a
random matrix interpretation. Indeed if we consider an ensemble of
random $N\times N$ Hermitian matrices $H$ with the joint
probability density of eigenvalues (see Mehta \cite{mehta1})
\begin{equation}\label{ProbabilityDistribution}
{\mathcal{P}}^{(N)}(x_1,\ldots ,x_N)=\frac{1}{Z_{N}}
\exp\left\{-N\sum\limits_{i=1}^{N}V(x_i)\right\}
\triangle^{2}(x_1,\ldots ,x_N)
\end{equation}
the monic orthogonal polynomial $\pi_N(x)$ can be understood as an
average of the \textit{characteristic polynomial}
${\mathcal{Z}}_N[x,H]= \mbox{det}(x-H)$ over the ensemble
\begin{equation}\label{PiZ}
\pi_N(x)=\left\langle {\mathcal{Z}}_N[x,H]\right\rangle_H
\end{equation}
A natural question which arises at this point is the following:
what do we obtain if instead of one characteristic polynomial we
average a combination of characteristic polynomials (for example,
their product). This question was recently addressed in Brezin and
Hikami \cite{brezin1}, Mehta and Normand \cite{mehta2} who derived
a generalization of formula (\ref{PiZ}). The authors considered
the ensemble average of products of characteristic polynomials. It
was found that this average is expressed as a determinant:
\begin{equation}\label{BrezinDeterminantPolynomials}
\left\langle \prod\limits_{j=1}^{L}{\mathcal{Z}}_N[\lambda_j,H]
\right\rangle_H=\frac{1}{\triangle(\hat\lambda)}\;
\mbox{det}\left|
\begin{array}{cccc}
  \pi_N(\lambda_1) & \pi_{N+1}(\lambda_1)
   & \ldots & \pi_{N+L-1}(\lambda_1) \\
  \pi_N(\lambda_2) & \pi_{N+1}(\lambda_2)
   & \ldots & \pi_{N+L-1}(\lambda_2) \\
  \vdots &  &  &  \\
  \pi_N(\lambda_L) & \pi_{N+1}(\lambda_L)
   & \ldots & \pi_{N+L-1}(\lambda_L)
\end{array}\right|
\end{equation}
where $\Delta(\hat{\lambda})\equiv
\Delta(\lambda_1,\ldots,\lambda_L)$. (See also a related paper by
Forrester and Witte \cite{For} which deals with a particular case
of positive integer moments: $\lambda_1=\ldots =\lambda_L$).

In the present paper we obtain a further generalization of
formulas (\ref{PiZ}) and (\ref{BrezinDeterminantPolynomials}).
Namely we compute the general spectral correlation function of
characteristic polynomials defined as
\begin{equation}\label{korrelationfunction}
{\mathcal{K}}_N(\hat\epsilon, \hat\mu)= \left\langle \frac{
\prod\limits_{j=1}^{L}{\mathcal{Z}}_N\left[\mu_j,
H\right]}{\prod\limits_{j=1}^{M}{\mathcal{Z}}_N\left[\epsilon_j,H\right]}
\right\rangle_{H}
\end{equation}
Here the symbols $\hat\epsilon ,\hat\mu$ denote the vectors with
the components:
\begin{eqnarray}
\hat\epsilon=(\epsilon_1,\ldots ,\epsilon_M);\qquad
\hat\mu=(\mu_1,\ldots ,\mu_L)
\end{eqnarray}
For the function ${\mathcal{K}}_N(\hat\epsilon, \hat\mu)$ to be
well defined we assume that $\mbox{Im}\;\epsilon_k\neq 0$, and
also assume $N\ge M, N\ge L$. Then the result we obtain for the
correlation function ${\mathcal{K}}_N(\hat\epsilon, \hat\mu)$ is
\begin{equation}\label{NewFormulaForKorrelation}
{\mathcal{K}}_N(\hat\epsilon, \hat\mu)=\frac{\prod_{j=N-M}^{N-1}
\gamma_j}{\triangle(\hat\mu)\triangle(\hat \epsilon)}\;\;
\mbox{det} \left|
\begin{array}{cccc}
  h_{N-M}(\epsilon_1) & h_{N-M+1}(\epsilon_1)
   & \ldots & h_{N+L-1}(\epsilon_1) \\
  \vdots &  &  &  \\
  h_{N-M}(\epsilon_M) & h_{N-M+1}(\epsilon_M)
   & \ldots & h_{N+L-1}(\epsilon_M) \\
  \pi_{N-M}(\mu_1) & \pi_{N-M+1}(\mu_1) & \ldots
   & \pi_{N+L-1}(\mu_1) \\
  \vdots &  &  &  \\
  \pi_{N-M}(\mu_L) & \pi_{N-M+1}(\mu_L)
   & \ldots & \pi_{N+L-1}(\mu_L)
\end{array}\right|
\end{equation}
where $h_k(\epsilon)$ denotes the Cauchy transform of the monic
orthogonal polynomial $\pi_k(x)$,
\begin{equation}\label{CauchyTransformOfMonic}
h_k(\epsilon)=\frac{1}{2\pi
i}\int\frac{\pi_k(x)e^{-NV(x)}dx}{x-\epsilon},
\;\;\;\mbox{Im}\;\epsilon\neq 0
\end{equation}
and
\begin{equation}\label{gamma}
\gamma_{n-1}=-\frac{2\pi i}{c^2_{n-1}}
\end{equation}
The formula above is of interest since it shows the place of the
correlation function ${\mathcal {K}}_N(\hat{\epsilon},\hat{\mu})$
in the theory of orthogonal polynomials. Furthermore it provides
new insight on ${\mathcal {K}}_N(\hat{\epsilon},\hat{\mu})$, for
example it relates ${\mathcal {K}}_N(\hat{\epsilon},\hat{\mu})$
with Riemann-Hilbert problem for orthogonal polynomials (see
section 5)

Correlation functions ${\mathcal {K}}_N(\hat{\epsilon},\hat{\mu})$
are interesting objects themselves  since they contain a very
detailed information about spectra of random matrices. In
particular, knowledge of such correlations enables to extract the
n-point correlation function of spectral densities. Moreover,
distributions of some interesting physical quantities such as,
e.g. level curvatures are expressed in terms of
${\mathcal{K}}_N(\hat{\epsilon} ;\hat{\mu})$, (see e.g \cite{AS}
and  Appendix A of \cite{Schom} for more examples).

The investigation of correlation functions of characteristic
polynomials and of their moments is also motivated by hope to
relate statistics of zeroes of the Riemann zeta function to that
of eigenvalues of large random matrices \cite{KS,brezin1}. The
spectral determinants are also relevant for interesting
combinatorial problems, see \cite{zhenya}. Other numerous
applications of spectral determinants are in the theory of quantum
chaotic and disordered systems, and in quantum chromodynamics, see
\cite{I}-\cite{II} for an extensive list of references.

There are several analytical techniques for dealing with the
integer moments (positive or negative) of characteristic
polynomials and more general correlations functions. Their
applicability varies with the nature of the underlying random
matrix ensemble. The particular case of the Gaussian measure -
Gaussian Unitary Ensemble (GUE), as well as its chiral
counterpart, chGUE - can be studied very efficiently by a
modification of the standard supersymmetry technique \cite{Efetov}
suggested recently by the authors \cite{I,II,chir}. One starts, as
usual, by representing each of the characteristic polynomials as a
Gaussian integral over anti-commuting (Grassmann) variables and
inverse characteristic polynomial by Gaussian integral over
complex variables. This allows
  to average the resulting expressions
straightforwardly. At the next step one employs the so-called
Hubbard-Stratonovich transformation for "fermionic" degrees of
freedom, and then
  exploits the Itzykson-Zuber-Harish-Chandra integrals
together with their natural non-compact extensions. A detailed
account of the method and results as well as related references
can be found in the recent papers \cite{I}-\cite{chir}.

The resulting expressions in \cite{II,chir} revealed a very
attractive determinantal structure which was especially evident in
the case of chiral GUE (Laguerre ensemble) \cite{chir}, see also
\cite{guhr}. When one deals with the positive moments only such
structures naturally arises in the framework of the orthogonal
polynomial method \cite{brezin1,mehta2}. This fact is suggestive
of the idea that the determinantal structure in the general case
(i.e. when both products and ratios of characteristic polynomials
are involved) should be valid for an arbitrary unitary-invariant
potential. In present paper we show that this is indeed the case.

The paper is organized as follows. In section 2 we derive an
algebraic identity which represents $\prod\limits_{j=1}^{M}
{\mathcal{Z}}_{N}^{-1}[\epsilon_j,H]$ as a sum over permutations.
This identity together with  formula
(\ref{BrezinDeterminantPolynomials}) will permit us to rewrite
${\mathcal{K}}_{N}(\hat\epsilon,\hat\mu)$ as an integral over $M$
variables with subsequent derivation of formula
(\ref{NewFormulaForKorrelation}).  Section 3  illustrates our
approach on the simplest case $M=1$ and $L=0$. Here we derive a
multi- variable integral representation for Cauchy transforms of
monic orthogonal polynomials. The computation of the general
correlation function ${\mathcal{K}}_{N}(\hat\epsilon,\hat\mu)$ for
any integers $0\leq L,M\leq N$ is given in section 4. In section 5
we establish relation with the Riemann-Hilbert problem for
orthogonal polynomials proposed by Fokas, Its and Kitaev
\cite{fokas1,fokas2} (see also \cite{univ1,deift1}), and outline a
way to investigate the large $N$ asymptotic of
${\mathcal{K}}_{N}(\hat\epsilon,\hat\mu)$ by Deift-Zhou steepest
descent/stationary phase method
\cite{deift1,deift2,deift3,deift4,deift5,deift6}. A detailed
asymptotic analysis of ${\mathcal{K}}_{N}(\hat\epsilon,\hat\mu)$
is beyond the scope of this paper and will be given in the
subsequent publication \cite{strahov1}. Finally in section 6 we
briefly discuss some "duality relations" for matrix integrals
emerging for the case of the Gaussian potential.

\section{Inverse of products of characteristic polynomials as a
sum over permutations} Let $M\leq N$, and $x_1,\ldots ,x_N$ denote
the eigenvalues of a matrix $H$ ($\mbox{dim}\;H=N\times N$). Then
the following algebraic identity  holds:
\begin{eqnarray}\label{AlgebraicIdentity}
\prod\limits_{l=1}^{M}\frac{\epsilon_l^{N-M}}{{\mathcal{Z}}_N[\epsilon_l,H]}=\qquad
\qquad\qquad\qquad\qquad\qquad\\
\sum\limits_{\sigma\in\; {\sf{S}}_N/{\sf{S}}_{N-M}\times
{\sf{S}}_M} \left( \frac{\prod\limits_{i=1}^M
x_{\sigma(i)}^{N-M}}{\prod\limits_{i,j=1}^M\epsilon_{j}-
x_{\sigma(i)}}\right)\; \frac{\triangle(x_{\sigma(1)},\ldots
,x_{\sigma(M)})\triangle(x_{\sigma(M+1)},\ldots
,x_{\sigma(N)})}{\triangle(x_{\sigma(1)},\ldots
,x_{\sigma(N)})}\nonumber
\end{eqnarray}
where ${\sf{S}}_N$ is the permutation group of the indices
$1,\ldots ,N$; ${\sf{S}}_M$ is the permutation group of the first
$M$ indices and ${\sf{S}}_{N-M}$ is the permutation group of the
remaining $N-M$ indices.  Identity (\ref{AlgebraicIdentity})
follows as a consequence of the Cauchy-Littlewood formula
\cite{schur}
\begin{equation}\label{Cauchy-Littlewood formula}
\prod\limits_{j=1}^{M}\prod\limits_{i=1}^{N}(1-x_iy_j)^{-1}=
\sum\limits_{\lambda}s_{\lambda}(x_1,\ldots
,x_N)s_{\lambda}(y_1,\ldots ,y_M)
\end{equation}
and the Jacobi-Trudi identity \cite{schur}:
\begin{equation}
s_{\lambda}(x_1,\ldots,x_N)=
\frac{\mbox{det}\left(x_i^{\lambda_j-j+N}\right)
}{\triangle(x_1,\ldots ,x_N)}
\end{equation}
where the Schur polynomial $s_{\lambda}(x_1,\ldots ,x_N)$
corresponds to the partition $\lambda$ and the indices $i,j$ take
the values from $1$ to $N$. In order to prove
(\ref{AlgebraicIdentity}) we rewrite the Schur polynomial
$s_{\lambda}(x_1,\ldots ,x_N)$ in (\ref{Cauchy-Littlewood
formula}) as
\begin{equation}\label{JacobiTrudi}
s_{\lambda}(x_1,\ldots ,x_N)=\frac{\sum\limits_{\pi\;\in
\;\textsf{S}_N}(-)^{\nu_{\pi}}x_{\pi(1)}^{\lambda_1-1+N}\ldots
x_{\pi(M)}^{\lambda_M-M+N}x_{\pi(M+1)}^{-M-1+N}\ldots
x_{\pi(N)}^{0}}{\triangle(x_1,\ldots ,x_N)}
\end{equation}
(As $s_{\lambda}(y_1,\ldots ,y_M)=0$ for any partition with the
number of rows larger then $M$ only the partitions with
$\lambda_{M+1}=\ldots =\lambda_N=\ldots =0$ contribute in equation
(\ref{Cauchy-Littlewood formula})). In equation
(\ref{JacobiTrudi}) $\nu_{\pi}=1$ (or $\nu_{\pi}=0$) if the
permutation $\pi\in\;\textsf{S}_N$ is odd (even).

Any permutation $\pi\in\textsf{S}_N$ in the above sum can be
decomposed as a product of three subsequent permutations, i.e.
$\pi=\sigma\cdot\pi_2\cdot\pi_1$. The first one, $\pi_1\in
\textsf{S}_{M}$ is a permutation  of elements of the set
$[1,2,\ldots ,M]$; the second permutation, $\pi_2\in
\textsf{S}_{N-M}$ is that of elements of the set $[M+1,M+2,\ldots
,N]$; and the third permutation, $\sigma\in
\textsf{S}_N/\textsf{S}_M\times \textsf{S}_{N-M}$, is an exchange
of elements between these two sets. For example, let us take
$M=3$, $N=5$, and permutation $\pi=\left(\begin{array}{ccccc}
  1 & 2 & 3 & 4 & 5 \\
  1 & 4 & 2 & 5 & 3
\end{array}
\right)$. This permutation can be represented as follows:
\begin{eqnarray}
\pi=\sigma\cdot\pi_2\cdot\pi_1\qquad\qquad\qquad\qquad\qquad\qquad\qquad\qquad \nonumber\\
\pi_1=\left(\begin{array}{ccccc}
  1 & 2 & 3 & 4 & 5 \\
  1 & 3 & 2 & 4 & 5
\end{array}
\right),\;\;\; \pi_2= \left(\begin{array}{ccccc}
  1 & 2 & 3 & 4 & 5 \\
  1 & 2 & 3 & 5 & 4
\end{array}
\right),\;\;\; \sigma=\left(\begin{array}{ccccc}
  1 & 2 & 3 & 4 & 5 \\
  1 & 5 & 3 & 4 & 2
\end{array}
\right) \nonumber
\end{eqnarray}
We now rewrite the denominator of the right-hand side of equation
(\ref{JacobiTrudi}) as
\begin{eqnarray}
\triangle(x_1,\ldots
,x_N)=(-)^{\nu_{\pi}}\triangle(x_{\pi(1)},\ldots ,x_{\pi(N)})=
\qquad\quad\quad\\
(-)^{\nu_{\pi}}\;\triangle(x_{\pi(1)},\ldots
,x_{\pi(M)})\;\triangle(x_{\pi(M+1)},\ldots
,x_{\pi(N)})\prod\limits_{i\in[1,\ldots ,M],\;j\in[M+1,\ldots ,
N]}(x_{\pi(i)}-x_{\pi(j)})\nonumber
\end{eqnarray}
and observe that
\begin{equation}
\sum\limits_{\pi_1\in\;
\textsf{S}_M}\frac{x_{\pi(1)}^{\lambda_1-1+N}\ldots
x_{\pi_1(M)}^{\lambda_M-M+N}}{\triangle(x_{\pi_1(1)},\ldots
,x_{\pi_1(M)})}=(x_1\ldots x_M)^{N-M}s_{\lambda}(x_1,\ldots ,x_M)
\end{equation}
and
\begin{equation}
\sum\limits_{\pi_2\in\;
\textsf{S}_{N-M}}\frac{x_{\pi_2(M+1)}^{-M-1+N}\ldots
x_{\pi_2(N)}^{0}}{\triangle(x_{\pi_2(M+1)},\ldots
,x_{\pi_2(N)})}=1
\end{equation}
After the permutations of type $\pi_1$ and $\pi_2$ have been
performed in equation ({\ref{JacobiTrudi}) we remain with the
following sum over permutations
$\sigma\in\;\textsf{S}_N/\textsf{S}_M\times \textsf{S}_{N-M}$:
\begin{eqnarray}\label{SchurExpression}
s_{\lambda}(x_1,\ldots ,x_N)=
\sum\limits_{\sigma\in\;\textsf{S}_N/\textsf{S}_M\times
\textsf{S}_{N-M}}\frac{\left(x_{\sigma(1)}^{N-M}\ldots
x_{\sigma(M)}^{N-M}\right)\;s_{\lambda}(x_{\sigma(1)}\ldots
x_{\sigma(M)})}{\prod\limits_{i\in[1,\ldots ,M],\;j\in[M+1,\ldots
, N]}(x_{\sigma(i)}-x_{\sigma(j)})}=\qquad\qquad\qquad \\
\sum\limits_{\sigma\in\;\textsf{S}_N/\textsf{S}_M\times
\textsf{S}_{N-M}}\frac{\left(\prod\limits_{j=1}^{M}x_{\sigma(j)}^{N-M}
\right)\;s_{\lambda}(x_{\sigma(1)}\ldots
x_{\sigma(M)})\triangle(x_{\sigma(1)}\ldots
x_{\sigma(M)})\triangle(x_{\sigma(M+1)}\ldots
x_{\sigma(N)})}{\triangle(x_{\sigma(1)}\ldots x_{\sigma(N)})}
\nonumber
\end{eqnarray}
We insert expression (\ref{SchurExpression}) to the sum in
equation (\ref{Cauchy-Littlewood formula}) and apply the
Cauchy-Littlewood formula. It gives
\begin{eqnarray}\label{identity}
\frac{1}{\prod\limits_{j=1}^{M}\prod\limits_{i=1}^{N}\left(1-x_iy_j\right)}=
\qquad\qquad\qquad\qquad\qquad\qquad\\
\sum\limits_{\sigma\in \textsf{S}_N/\textsf{S}_{N-M}\times
\textsf{S}_M}\left( \frac{\prod\limits_{i}^{M}x_{\sigma(i)}
^{N-M}}{\prod\limits_{i,j=1}^{M}\left(1-x_{\sigma(i)}y_j\right)
}\right)\frac{\triangle\left(x_{\sigma(1)},\ldots
,x_{\sigma(M)}\right)\triangle\left(x_{\sigma(M+1)},\ldots
,x_{\sigma(N)}\right)}{\triangle\left(x_{\sigma(1)},\ldots
,x_{\sigma(N)}\right)}\nonumber
\end{eqnarray}
It is easy to see that the formula above is equivalent to the
formula (\ref{AlgebraicIdentity}).
\section{Multi-variable integral representation for Cauchy transforms
of orthogonal polynomials} In this section we compute
$\left\langle{\mathcal{Z}}_N^{-1}[\epsilon,H]\right\rangle_H $,
$\mbox{dim} H=N$, and show that this average is equal to
$\gamma_{N-1}h_{N-1}(\epsilon)$. Here $\gamma_{N-1}$ is given by
equation (\ref{gamma}) and $h_{N-1}(\epsilon)$ is the Cauchy
transform of the monic polynomial $\pi_{N-1}(x)$ (equation
(\ref{CauchyTransformOfMonic})). The aim is just to illustrate our
approach to the computation of the general correlation function
${\mathcal{K}}_N(\hat\epsilon,\hat\mu)$ of characteristic
polynomials on the simple example $L=0, M=1$.

Once $M=1$  algebraic identity (\ref{AlgebraicIdentity}) takes the
form:
\begin{equation}
\frac{\epsilon^{N-M}}{{\mathcal{Z}}_{N}[\epsilon,H]}=
\sum\limits_{\sigma\;\in\;\textsf{S}_N/\textsf{S}_{N-1}}
\left(\frac{x_{\sigma(1)}^{N-1}}{\epsilon-x_{\sigma(1)}}
\right)\frac{\triangle(x_{\sigma(2)},\ldots
,x_{\sigma(N)})}{\triangle(x_{\sigma(1)},\ldots ,x_{\sigma(N)})}
\end{equation}
Averaging the expression above with the probability measure
${\mathcal{P}}^{(N)}(x_1,\ldots ,x_N)$ (equation
(\ref{ProbabilityDistribution})) we notice that each term yields
the same contribution (${\mathcal{P}}^{(N)}(x_1,\ldots ,x_N)$ is
symmetric with respect to the permutations). Thus we have
\begin{eqnarray}
\left\langle{\mathcal{Z}}_N^{-1}[\epsilon,H]\right\rangle_H =
\qquad\qquad\qquad\nonumber\\
\frac{\epsilon^{M-N}N}{Z_N}\int\frac{x_1^{N-1}}{\epsilon-x_1}
\triangle(x_2,\ldots ,x_N)\triangle(x_1,\ldots ,x_N) e^{-N\sum
\limits_{i=1}^N V(x_i)} dx_1\ldots dx_N
\end{eqnarray}
The integral above can be further rewritten as
\begin{equation}
Z_{N-1}\int dx_1\frac{x_1^{N-1}e^{-NV(x_1)}}{\epsilon-x_1}
\left\langle{\mathcal{Z}}_{N-1}[x_1,\tilde H]\right\rangle_{\tilde
H},\;\;\;\;\mbox{dim}\; \tilde H=N-1\nonumber
\end{equation}
Using the formula (\ref{PiZ}) we then obtain
\begin{equation}
\left\langle{\mathcal{Z}}_N^{-1}[\epsilon,H]\right\rangle_H =
N\epsilon^{N-M}\frac{Z_{N-1}}{Z_N}\int
\frac{x^{N-1}\pi_{N-1}(x)e^{-NV(x)}dx}{\epsilon-x}
\end{equation}
Now we note that
\begin{equation}
\frac{x^{N-1}}{\epsilon-x}=\frac{x^{N-1}-\epsilon^{N-1}}{\epsilon-x}+
\frac{\epsilon^{N-1}}{\epsilon-x}\nonumber
\end{equation}
The first term above is a polynomial of degree $N-2$ which is
orthogonal to the polynomial$\pi_{N-1}(x)$. Thus  only the second
term contributes to the integral. It gives
\begin{equation}
\left\langle{\mathcal{Z}}_N^{-1}[\epsilon,H]\right\rangle_H =
\gamma_{N-1}h_{N-1}(\epsilon)
\end{equation}
or
\begin{equation}\label{IntegralRepresentofCauchyTransform}
\gamma_{N-1}h_{N-1}(\epsilon)=\frac{1}{Z_N}\int\prod\limits_{j=1}^{N}(\epsilon
-x_j)^{-1}\;e^{-N\sum\limits_{j=1}^{N}V(x_j)}\triangle^2(x_1,\ldots
, x_N)dx_1\ldots x_N
\end{equation}
where we have used formulas (\ref{CauchyTransformOfMonic}),
(\ref{gamma}) and have expressed $Z_N$ as
$N!\prod\limits_{j=0}^{N-1}c_j^2$.

We expect that the representation of Cauchy transforms of
orthogonal polynomials as multi-variable integrals
(\ref{IntegralRepresentofCauchyTransform}) might be known, but we
failed to trace it in the standard monographs on the subject.

\section{Derivation of the formula for
${\mathcal{K}}_N(\hat\epsilon,\hat\mu)$} Let us now consider the
general case. The correlation function
${\mathcal{K}}_N(\hat\epsilon,\hat\mu)$ defined by equation
(\ref{korrelationfunction}) is an integral over $N$ variables
$x_1,\ldots ,x_N$ (which are eigenvalues of $H$ ) with an
integrand symmetric under permutations. It then follows that each
component of the sum in equation (\ref{AlgebraicIdentity}) gives
the same contribution to the correlation function
${\mathcal{K}}_N(\hat\epsilon,\hat\mu)$. Thus we have
\begin{eqnarray}
{\mathcal{K}}_N(\hat\epsilon,\hat\mu)=\frac{N!}{(N-M)!M!}
\prod\limits_{j=1}^{M}\epsilon_j^{M-N} \int dx_1\ldots dx_N
{\mathcal{P}}^{(N)}
(x_1,\ldots ,x_N)\nonumber\\
\times\left(\prod\limits_{i,j=1}^{M}\frac{x_i^{N-M}}{\epsilon_j-x_i}\right)
\frac{\prod\limits_{j=1}^{L}{\mathcal{Z}}_N[\mu_j,H]}{
\prod\limits_{i=1}^{M}\prod\limits_{j=M+1}^{N}(x_i-x_j)}
\end{eqnarray}
We decompose the eigenvalue probability density function as
\begin{eqnarray}
{\mathcal{P}}^{(N)}(x_1,\ldots
,x_N)=\frac{Z_{N-M}Z_M}{Z_N}\prod\limits_{i=1}^{M}\prod\limits_{j=M+1}^{N}
(x_i-x_j)^2\\
\times\; {\mathcal{P}}^{(M)}(x_1,\ldots
,x_M){\mathcal{P}}^{(N-M)}(x_{M+1},\ldots ,x_{N})\nonumber
\end{eqnarray}
With the decomposition above the integral expression for the
correlation function ${\mathcal{K}}_N(\hat\epsilon,\hat\mu)$ can
be rewritten as
\begin{eqnarray}\label{KasMFoldIntegral}
{\mathcal{K}}_N(\hat\epsilon,\hat\mu)=\frac{N!}{(N-M)!M!}
\left(\prod\limits_{j=1}^{M}\epsilon_j^{M-
N}\right)\frac{Z_{N-M}Z_{M}}{Z_N}\quad\qquad\\
\int dx_1\ldots dx_M {\mathcal{P}}^{(M)}(x_1,\ldots
,x_M)F_{\hat\epsilon,\hat\mu}(x_1,\ldots
,x_M)I^{(N-M)}(x_1,\ldots,x_M,\hat\mu)\nonumber
\end{eqnarray}
where we have introduced two functions of $M$ variables
$x_1,\ldots ,x_M$. The first function is
\begin{equation}
F_{\hat\epsilon,\hat\mu}(x_1,\ldots ,x_M)=
\prod\limits_{i,j=1}^{M}\prod\limits_{k=1}^L
\frac{x_i^{N-M}(\mu_k-x_i)}{\epsilon_j-x_i}
\end{equation}
The second function, $I^{(N-M)}(x_1,\ldots,x_M,\hat\mu)$, can be
understood as an averaged value of the products of characteristic
polynomials over an ensemble of $N-M$ dimensional Hermitian
matrices $\tilde{H}$ with eigenvalues $x_{M+1},\ldots ,x_N$,
\begin{eqnarray}
I^{(N-M)}(x_1,\ldots,x_M,\hat\mu)=\qquad\qquad\qquad\\
\left\langle \prod\limits_{i=1}^M \prod\limits_{k=1}^{L}
{\mathcal{Z}}_{N-M}[x_i,\tilde{H}]
{\mathcal{Z}}_{N-M}[\mu_k,\tilde{H}]\nonumber
\right\rangle_{\tilde{H}}
\end{eqnarray}
It is convenient to introduce $L+M$-dimensional vector $\hat r$
whose components are the integration variables $x_1,\ldots ,x_M$
and the elements of the vector  $\hat\mu$, i.e.
\begin{equation}
\hat r=(x_1,\ldots ,x_M,\mu_1,\ldots ,\mu_L)
\end{equation}
We exploit  formula (\ref{BrezinDeterminantPolynomials}) which for
$I^{(N-M)}(x_1,\ldots,x_M,\hat\mu)\equiv I^{(N-M)}(\hat r) $ gives
the following expression
\begin{equation}\label{IDeterminant}
I^{(N-M)}(\hat r)=\frac{1}{\triangle(\hat r)}\;
\mbox{det}\left[\pi_{N-M+j-1}(r_i)\right]_{1\leq i,j\leq L+M}
\end{equation}
The Vandermonde determinant $\triangle(\hat r)$ can be factorized
as
\begin{equation}\label{VandermondeFactorization}
\triangle(\hat r)=\left\{
\prod\limits_{k=1}^{L}\prod\limits_{j=1}^{M} (x_j-\mu_k) \right\}
\triangle(x_1,\ldots ,x_M)\triangle(\mu_1,\ldots ,\mu_k)
\end{equation}
Now we insert equations (\ref{IDeterminant}) and
(\ref{VandermondeFactorization}) to the formula for the
correlation function ${\mathcal{K}}_{N}(\hat\epsilon,\hat\mu)$
(equation (\ref{KasMFoldIntegral})). It gives:
\begin{equation}\label{KasIntegralOfDeterminants}
{\mathcal{K}}_N(\hat\epsilon ,\hat\mu)=\frac{N!}{(N-M)!M!}
\left(\prod\limits_{j=1}^{M}\epsilon_j^{M-N}\right)\frac{Z_{N-M}}{Z_N}
\frac{1}{\triangle(\hat\lambda,
\hat\mu)}J_{M}(\hat\epsilon,\hat\mu)
\end{equation}
where $J_{M}(\hat\epsilon,\hat\mu)$ is the $M$- fold integral
\begin{eqnarray}\label{IntegralJM1}
J_{M}(\hat\epsilon,\hat\mu)=\qquad\qquad\qquad\\\int dx_1\ldots
dx_m \;\mbox{det}\left[\Phi_i(x_j)\right]_{1\leq i,j\leq M}
\mbox{det}\left[\pi_{N-M+j-1}(r_i)\right]_{1\leq i,j\leq
L+M}\nonumber
\end{eqnarray}
where we have introduced $M$ functions $\Phi_i(x)$
\begin{equation}
\Phi_i(x)=\frac{x^{i-1}G(x)}{\prod\limits_{j=1}^{M}(\epsilon_j-x)},\;\;
G(x)=e^{-NV(x)}x^{N-M}
\end{equation}
Thus  the computation of the correlation function
${\mathcal{K}}_N(\hat\epsilon,\hat\mu)$ is reduced to that of the
integral of $M$ variables. This integral can be further
transformed to a determinant form. To proceed we first note that
$\mbox{det}\left[\Phi_i(x_j)\right]$ can be simplified\footnote{
This follows from the algebraic relation \\
\begin{equation}
\frac{x_j^{i-1}G(x_j)}{\prod\limits_{l=1}^{M}(\epsilon_l-x_j)}=
(-)^M\sum\limits_{K=1}^{M}\frac{\epsilon_k^{i-1}}{\prod\limits_{l\neq
k}(\epsilon_k-\epsilon_l)}\;\frac{G(x_j)}{x_j-\epsilon_k}\nonumber
\end{equation} \\
which can be derived from identity (\ref{AlgebraicIdentity}). With
the equation above we have
\begin{equation}
\mbox{det}\left[\Phi_i(x_j)\right]=(-)^{M^2}
\mbox{det}\left(\frac{G(x_i)}{\epsilon_j-x_i}\right) \mbox{det}
\left(\epsilon^{i-1}_j\prod\limits_{l\neq
j}\frac{1}{\epsilon_j-\epsilon_l}\right)\nonumber
\end{equation}
Noting that
\begin{equation}
(-)^{M^2}\mbox{det} \left(\epsilon^{i-1}_j\prod\limits_{l\neq
j}\frac{1}{\epsilon_j-\epsilon_l}\right) =
\frac{1}{\triangle(\hat\epsilon)}\nonumber
\end{equation}
we obtain equation (\ref{determinantsimplification}) }
\begin{equation}\label{determinantsimplification}
\mbox{det}\left[\Phi_i(x_j)\right]=\frac{1}{\triangle(\hat\epsilon)}
\;\mbox{det}\left(\frac{G(x_i)}{\epsilon_j-x_i}\right) \, .
\end{equation}
Insert the expression (\ref{determinantsimplification}) to the
integral (\ref{IntegralJM1}), rewrite the determinants as sums
over permutations and find in this way that
\begin{equation}\label{determinantJ1M}
J_{M}(\hat\epsilon,\hat\mu)=
\frac{M!}{\triangle(\hat\epsilon)\triangle(\hat\mu)}\;\;
\mbox{det} \left|
\begin{array}{ccc}
  \tilde{\pi}_{N-M}(\epsilon_1) & \ldots & \tilde{\pi}_{N+L-1}(\epsilon_1)  \\
  \vdots &  &  \\
  \tilde{\pi}_{N-M}(\epsilon_M) & \ldots & \tilde{\pi}_{N+L-1}(\epsilon_M)  \\
  \pi_{N-M}(\mu_1) & \ldots & \pi_{N+L-1}(\mu_1)  \\
  \vdots &  &  \\
  \pi_{N-M}(\mu_L) & \ldots & \pi_{N+L-1}(\mu_L)  \\
\end{array}
\right|
\end{equation}
where
\begin{equation}
\tilde{\pi}_k(\epsilon)=\int
\frac{e^{-NV(x)}x^{N-M}}{\epsilon-x}\pi_k(x)dx,\;\;\; k\in
[N-M,N-M+L-1] \, .
\end{equation}
Using the orthogonality of monic polynomials $\pi_k(x)$ with
respect to the measure $e^{-NV(x)}dx $ we observe the relation
between $\tilde{\pi}_k(\epsilon)$ and the Cauchy transforms
$h_k(\epsilon)$:
\begin{equation}\label{TildeRelation}
\tilde{\pi}_k(\epsilon)=-2\pi i\epsilon^{N-M}h_k(\epsilon)
\end{equation}
We insert expressions (\ref{determinantJ1M}),
(\ref{TildeRelation}) to formula (\ref{KasIntegralOfDeterminants})
for the correlation function
${\mathcal{K}}_{N}(\hat\epsilon,\hat\mu)$. Using the relations
between coefficients $Z_j$, $c_j$, $\gamma_j$  we finally prove
(\ref{NewFormulaForKorrelation}).
\section{Correlation function ${\mathcal{K}}_{N}(\hat\epsilon,\hat\mu)$
and the Riemann-Hilbert problem for orthogonal polynomials} Our
formula (\ref{NewFormulaForKorrelation}) enables us to express the
correlation function ${\mathcal{K}}_{N}(\hat\epsilon,\hat\mu)$ in
terms of solutions of Riemann-Hilbert problems for orthogonal
polynomials proposed by Fokas, Its and Kitaev
\cite{fokas1,fokas2}. It follows from
(\ref{NewFormulaForKorrelation}) that the correlation function
${\mathcal{K}}_{N}(\hat\epsilon,\hat\mu)$ is determined by monic
orthogonal polynomials and their Cauchy transforms.  In turn, the
monic orthogonal polynomials and their Cauchy transforms can be
understood as elements of a (matrix valued) solution of the
following Riemann-Hilbert problem. Let contour $\Sigma$ be the
real line oriented from left to right. The upper side of the
complex plane with respect to the contour will be called the
positive one and the down side- the negative one. Once an integer
$n\geq 0$ is fixed the Riemann-Hilbert problem is to find a
$2\times 2$ matrix valued function $Y=Y^{(n)}(z)$ satisfying the
following conditions:
\begin{itemize}
  \item $Y^{(n)}(z)- \mbox{analytic}\;\mbox{in}\;
   \textsc{C}\setminus\Sigma $
  \item $Y^{(n)}_{+}(z)=Y^{(n)}_{-}(z)\left(
  \begin{array}{cc}
    1 & e^{-nV(z)} \\
    0 & 1 \
  \end{array}
  \right),\;z\in \Sigma$
  \item $Y^{(n)}(z)\mapsto\left(I+{\mathcal{O}}(z^{-1})\right)
  \left(\begin{array}{cc}
    z^n & 0 \\
    0 & z^{-n} \
  \end{array}
  \right)\;\; \mbox{as}\;\; z\mapsto \infty $
\end{itemize}
Here  $Y^{(n)}_{\pm}(z)$ denotes the limit of  $Y^{(n)}(z')$ as
$z'\mapsto z\in \Sigma$ from the positive/negative side. As is
proved by Fokas, Its and Kitaev \cite{fokas1,fokas2} the solution
of the Riemann-Hilbert problem is unique and is given by
\begin{equation}\label{R-H Solution}
Y^{(n)}(z)=\left(
\begin{array}{cc}
  \pi_n(z) & h_n(z) \\
  \gamma_{n-1}\pi_{n-1}(z) & \gamma_{n-1}h_{n-1}(z)
\end{array}
\right),\;\;\;\mbox{Im}\;z\neq 0
\end{equation}
On comparing the formulae (\ref{NewFormulaForKorrelation}) and
(\ref{R-H Solution}) we observe that the correlation function
${\mathcal{K}}_{N}(\hat\epsilon,\hat\mu)$ can be expressed in
terms of the matrix elements of the solution of the above
Riemann-Hilbert problem. The relation provides us with a
possibility to investigate the asymptotics of
${\mathcal{K}}_{N}(\hat\epsilon,\hat\mu)$ at large $N$ for
essentially any potential function $V(x)$ entering the probability
distribution (\ref{ProbabilityDistribution}). The details will be
presented in a forthcoming publication \cite{strahov1}. Here we
just outline the main steps. First we will show that the formula
(\ref{NewFormulaForKorrelation}) can be rewritten as a determinant
whose entries are two kernel functions. Those kernel functions are
expressible in terms of the solution of the Riemann-Hilbert
problem for orthogonal polynomials. The large $N$ asymptotics in
the Dyson scaling limit can be studied by the steepest
descent/stationary phase method for Riemann-Hilbert problems
introduced by Deift and Zhou \cite{deift2} and developed further
in \cite{deift3,deift4,deift5,deift6}. As a result we will prove
the universality of various quantities related to
${\mathcal{K}}_{N}(\hat\epsilon,\hat\mu)$. In particular we shall
be able to prove the universality  of the distributions of the
level curvatures and the local density of states, as well as to
derive the Poisson kernel distribution of S-matrix in a random
matrix model of quantum chaotic scattering. Understanding of
universality of objects arising in random matrix theory in various
scaling limits was recently a subject of intensive work both in
physical \cite{univphys} as well as mathematical
\cite{univpast,univ1} communities.

\section{"Duality relations" for the Gaussian case}

For the particular case of the Gaussian measure $V(x)=x^2/2$
equation (\ref{BrezinDeterminantPolynomials}) for the ensemble
average of products of characteristic polynomials can be written
in an equivalent form:
\begin{equation}\label{duality}
\int_{N\times N} d[\hat{H}]e^{-\frac{N}{2}\mbox{\small
Tr}\hat{H}^2} \det\left(\hat{\lambda}\otimes {\bf 1}_N-{\bf
1}_L\otimes \hat{H}\right) \propto \int_{L\times L}
[d\hat{Q}]e^{-\frac{N}{2}\mbox{\small Tr}\hat{Q}^2}
\det{\left(\hat{\lambda}-i\hat{Q}\right)}^N
\end{equation}
Here the integration in the left-hand side goes over the manifold
of Hermitian matrices of the size $N\times N$, whereas in the
right-hand side it goes over $L\times L$ Hermitian matrices.

This integral identity can be viewed as a certain "duality"
relation for matrix integrals. It emerged in various physical
contexts, most notably in the context of matrix models of the
string theory (see e.g. \cite{Morozov}, p. 27) where it played an
important role in understanding equivalence between one-matrix
models of the quantum gravity and the so-called Kontsevich-type
models \cite{Morozov,Kharchev}. To understand equation
(\ref{duality}) one recalls that the orthogonal polynomials for
the Gaussian case are Hermite polynomials possessing an integral
representation:
\begin{equation}\label{intrep}
\pi_{k}(\lambda)\propto e^{N\lambda^2/2}\int_{-\infty}^{\infty}dq
q^{k} e^{-N\left(\frac{q^2}{2}-i\lambda q\right)}
\end{equation}
Substituting (\ref{intrep}) into
(\ref{BrezinDeterminantPolynomials}) one easily brings the
right-hand side of the latter to the form
\begin{eqnarray}\label{diag1}
\left\langle \prod\limits_{j=1}^{L}{\mathcal{Z}}_N[\lambda_j,H]
\right\rangle_H\propto \frac{e^{\frac{N}{2}\mbox{\small
Tr}\hat{\lambda}^2}} {\Delta\{\lambda\}}\int
d\hat{Q}_{\lambda}\Delta\{\hat{Q}_{\lambda}\}
e^{-N\left[\frac{1}{2}\mbox{\small
Tr}\hat{Q}_{\lambda}^2-i\mbox{\small Tr}
\hat{Q}_{\lambda}\hat{\lambda}\right]}\det{\hat{Q}}_{\lambda}^{N}
\end{eqnarray}
where $\hat{Q}_{\lambda}=\mbox{diag}(q_1,\ldots,q_{L})$. From the
other hand the expression above can be obtained after shifting
$\lambda-i\hat{Q}\to -i\hat{Q}$ in equation (\ref{duality}),
diagonalizing the $L\times L$ matrix $\hat{Q}$  and integrating
out the angular degrees of freedom with help of the
Itzykson-Zuber-Harish-Chandra formula. A particular case of the
formula (\ref{diag1}) was mentioned in \cite{For} in the context
of symmetric Jack polynomials.

It is easy to check that the Cauchy transforms of the Hermite
polynomials are given by
\begin{equation}\label{intrep1}
h_k(\epsilon) \propto \int dq q^{k}
e^{-N\left(\frac{q^2}{2}-i\mbox{\small sgn}(\mbox{\small Im}
\,\epsilon)\epsilon q\right)}
\end{equation}
Here the integration domain is $0< q <\infty$ for
$\mbox{Im}\;{\epsilon}>0$ and $-\infty< q <0$ for
$\mbox{Im}\;{\epsilon}<0$. This gives us a possibility to rewrite
our main object -  correlation function
(\ref{NewFormulaForKorrelation}) - in the form of $M+L$-fold
integral \cite{note}:
\begin{eqnarray}\label{diag}
{\cal K}_{N}(\hat{\epsilon},\hat{\mu})\propto
\frac{e^{\frac{N}{2}\mbox{\small Tr}\hat{\mu}^2}}
{\Delta\{\mu\}\Delta\{\epsilon\}} \int d\hat{Q}\Delta\{\hat{Q}\}
e^{-N\left[\frac{1}{2}\mbox{\small Tr}\hat{Q}^2-i\mbox{\small Tr}
\hat{Q}\hat{E}\right]}\det{\hat{Q}}^{N-M}
\end{eqnarray}
where $\hat{Q}=(q_1,\ldots,q_{L},q_{L+1},\ldots,q_{M+L})$ and
$\hat{E}=(\hat{\mu},\hat{\epsilon}_{+}, -\hat{\epsilon}_{-})$,
with $\epsilon_{+}$ and $\epsilon_{-}$ denoting spectral
parameters $\epsilon_k$ with positive (negative) imaginary part,
respectively. The integration domain is $-\infty<q_l<\infty$ for
$1\le l\le L$ but $0< q_l <\infty$ or $-\infty< q_l <0$ for
$L<l\le M+L$, depending on the sign of the imaginary part of the
corresponding spectral parameter.

Expression (\ref{diag}) is a generalization of  duality relation
(\ref{diag1}). It arises most naturally in the method based on
Gaussian integral representations and Itzykson-Zuber type
integrations \cite{II}. Similar identities hold for the chiral
ensemble (Laguerre polynomials), see \cite{chir} for more details.

\section{Conclusions}
In this paper we have found an exact formula for the general
correlation function  ${\mathcal{K}}_{N}(\hat\epsilon,\hat\mu)$
containing both products and ratios of characteristic polynomials.
Our result is valid for an arbitrary ensemble of Hermitian
matrices of $\beta=2$ class. The obtained formula  establishes a
correspondence between the correlation function
${\mathcal{K}}_{N}(\hat\epsilon,\hat\mu)$ and the Riemann-Hilbert
problem for orthogonal polynomials. It is remarkable as it enables
us to study  the large $N$ asymptotics of this correlation
function via the Riemann-Hilbert approach. Among interesting
prospects for future research we would like to mention a
challenging problem of extending our calculations to other
symmetry classes $\beta=1,4$ (cf. \cite{KV}) as well as to
ensembles of non-Hermitian random matrices important for the
problems of quantum chaotic scattering \cite{FK}.
\section{Acknowledgements}
We would like to thank A. Its, J. Keating, B. A. Khoruzhenko for
valuable comments and discussions. This research was supported by
EPSRC grant GR/13838/01 "Random matrices close to unitarity or
Hermitian".

\end{document}